\documentclass[12pt,a4paper]{article}
\usepackage{graphicx,fancyhdr}
\usepackage{cite,citesort}
\usepackage{color}
\usepackage{wasysym}
\begin{document}

\pagenumbering{arabic}

\vspace{-40mm}

\begin{center}
\Large\textbf{Fabrication of FeSi and Fe$_3$Si compounds by electron beam induced mixing of [Fe/Si]$_2$ and [Fe$_3$/Si]$_2$ multilayers grown by focused electron beam induced deposition}\normalsize \vspace{5mm}

\large F. Porrati${^1}$, R. Sachser${^1}$, G. C. Gazzadi${^2}$, S. Frabboni${^{2,3}}$ and M. Huth${^1}$

\vspace{5mm}

\small\textit{1. Physikalisches Institut, Goethe-Universit\"at, Max-von-Laue-Str.~1, D-60438 Frankfurt am Main, Germany}

\small\textit{2. S3 Center, Nanoscience Institute-CNR, Via Campi 213/a, 41125 Modena, Italy}

\small\textit{3. FIM Department, University of Modena and Reggio Emilia, Via G. Campi 213/a, 41125 Modena, Italy}

\newpage

\end{center}

\vspace{0cm}
\begin{center}
\large\textbf{Abstract}\normalsize
\end{center}

\hspace{-6mm}$\textit{Fe-Si}$ binary compounds have been fabricated by focused electron beam induced deposition by the alternating use of iron pentacarbonyl, $Fe(CO)_5$, and neopentasilane, $Si_5H_ {12}$ as precursor gases. The fabrication procedure consisted in preparing multilayer structures which were treated by low-energy electron irradiation and annealing to induce atomic species intermixing. In this way we are able to fabricate $FeSi$ and $Fe_3Si$ binary compounds from $[Fe/Si]_2$ and $[Fe_3/Si]_2$ multilayers, as shown by transmission electron microscopy investigations. This fabrication procedure is useful to obtain nanostructured binary alloys from precursors which compete for adsorption sites during growth and, therefore, cannot be used simultaneously.

\newpage

\begin{center}
\large\textbf{1. Introduction}\normalsize
\end{center}

\hspace{-6mm}$\textit{Fe-Si}$ binary alloys form a group of intermetallic compounds interesting for magnetic, optical and thermoelectric applications\cite{okamoto00,liang11}. For example, $\beta$-$FeSi_2$ is a semiconductor material with a band-gap of about 0.87~eV suitable for optoelectronic technology~\cite{leong97}. $FeSi$ is a narrow gap semiconductor with cubic B20 structure and unusual electronic and magnetic properties\cite{schlesinger93}. $Fe_3Si$ and $Fe_5Si_3$ are room-temperature ferromagnets attractive for spintronics \cite{herfort03,veetil10}. In particular, $Fe_3Si$ is a pseudo-Heusler alloy with $D0_3$ structure, equivalent to the $L2_1$ structure of full-Heusler compounds, which are predicted to be half-metallic~\cite{degroot88,felser07}. Conventionally, the preparation of $\textit{Fe-Si}$ binary alloys takes place by means of solid-state reactions~\cite{walser76,gaffet93}. Alternatively, organometallic precursors were used to fabricate $Fe_3Si$ and $Fe_5Si_3$ nanoparticles by pyrolysis~\cite{veetil10} and iron silicides by chemical vapor deposition(CVD)~\cite{dormans91}. CVD was also employed to grow polycrystalline $FeSi$ thin films and freestanding $FeSi$ nanowires from the organometallic single source precursor $Fe(SiCl_3)_2(CO)_4$~\cite{zybill94,schmitt10}.

Focused electron beam induced deposition (FEBID) allows the growth of binary alloys by means of organometallic precursors~\cite{che05,winhold11,CoPt,CoSi}. In FEBID the adsorbed molecules of a precursor gas injected in a scanning electron microscope (SEM) dissociate by the interaction with the electron beam depositing the sample during the rastering process~\cite{utke08,huth12}. Mainly, FEBID binary nanostructures have been prepared by the simultaneous use of two precursor gases, as in the case of $FePt$, $PtSi$, $CoPt$ and $CoSi$~\cite{che05,winhold11,CoPt,CoSi}. Recently, $CoFe$ nanostructures have been fabricated by means of a single source heteronuclear precursor~\cite{CoFe}. In this case, the stoichiometry of the alloy is identical to the one contained in the precursor gas. Although the use of single-source precursors greatly simplifies the fabrication process of binary nanostructures, the simultaneous use of two precursor gases is still of advantage to target a series of alloys with variable stoichiometry~\cite{CoSi}. However, if the two precursors compete for adsorption site, the deposition process is dominated by the precursor provided at a higher partial pressure. In this case binary alloys cannot be easily fabricated and alternative fabrication strategies have to be developed.

In this work, we fabricate $\textit{Fe-Si}$ binary compounds by using iron pentacarbonyl, $Fe(CO)_5$, and neopentasilane, $Si_5H_{12}$, as precursor gases. Our findings suggest  that these precursors compete for adsorption sites and, therefore, cannot be used simultaneously for the fabrication of binary systems. In order to overcome this problem, we grow $Fe/Si$ multilayers by alternating deposition using $Fe(CO)_5$ and  $Si_5H_{12}$, followed by treatment by low energy electron irradiation and annealing to induce atomic species intermixing. In this way we are able to obtain the desired binary compounds.

\begin{center}
\large\textbf{2. Experimental section}\normalsize
\end{center}

The samples were prepared by using a dual beam SEM/FIB microscope (FEI, Nova Nanolab~600), equipped with a Schottky electron emitter. A self-made coaxial gas injection system (GIS)~\cite{lukas} was employed to introduce into the SEM the $Fe(CO)_5$ and $Si_5H_{12}$ precursors via a capillary of 0.5~mm inner diameter. The distance capillary-substrate surface was about 1.7~mm. A number of samples were prepared by using $Fe(CO)_5$ and $Si_5H_{12}$ either simultaneously or alternately. The samples were grown on 100~nm thick $Si_3N_4$ membranes by using electron beam parameters of 5~keV, 1~nA, 20~nm, and 1~$\mu$s for beam energy, beam current, pitch, and dwell time, respectively. After fabrication the samples prepared by the alternating use of $Fe(CO)_5$ and $Si_5H_{12}$ were irradiated in-situ, without breaking the vacuum, with the electron beam at an energy of 5~keV, a beam current of 1~nA and a dose of 1~$\mu C/\mu m^2$ at 280$^{\circ}$~C to induce atomic species intermixing. Additionally, some of these samples, see details below, were annealed in a vacuum furnace at a base pressure of about $1\cdot10^{-5}$~mbar. In this case, the temperature was increased from room temperature to 600$^{\circ}$~C in four hours and then decreased to room temperature over night. After fabrication and post-growth treatment, the structure of the samples was investigated by transmission electron microscopy (TEM) using a JEOL 2010 at 200~keV.

\begin{center}
\large\textbf{3. Results}\normalsize
\end{center}

\begin{figure}\vspace{20mm}\hspace{0mm}{\includegraphics[width=11cm]{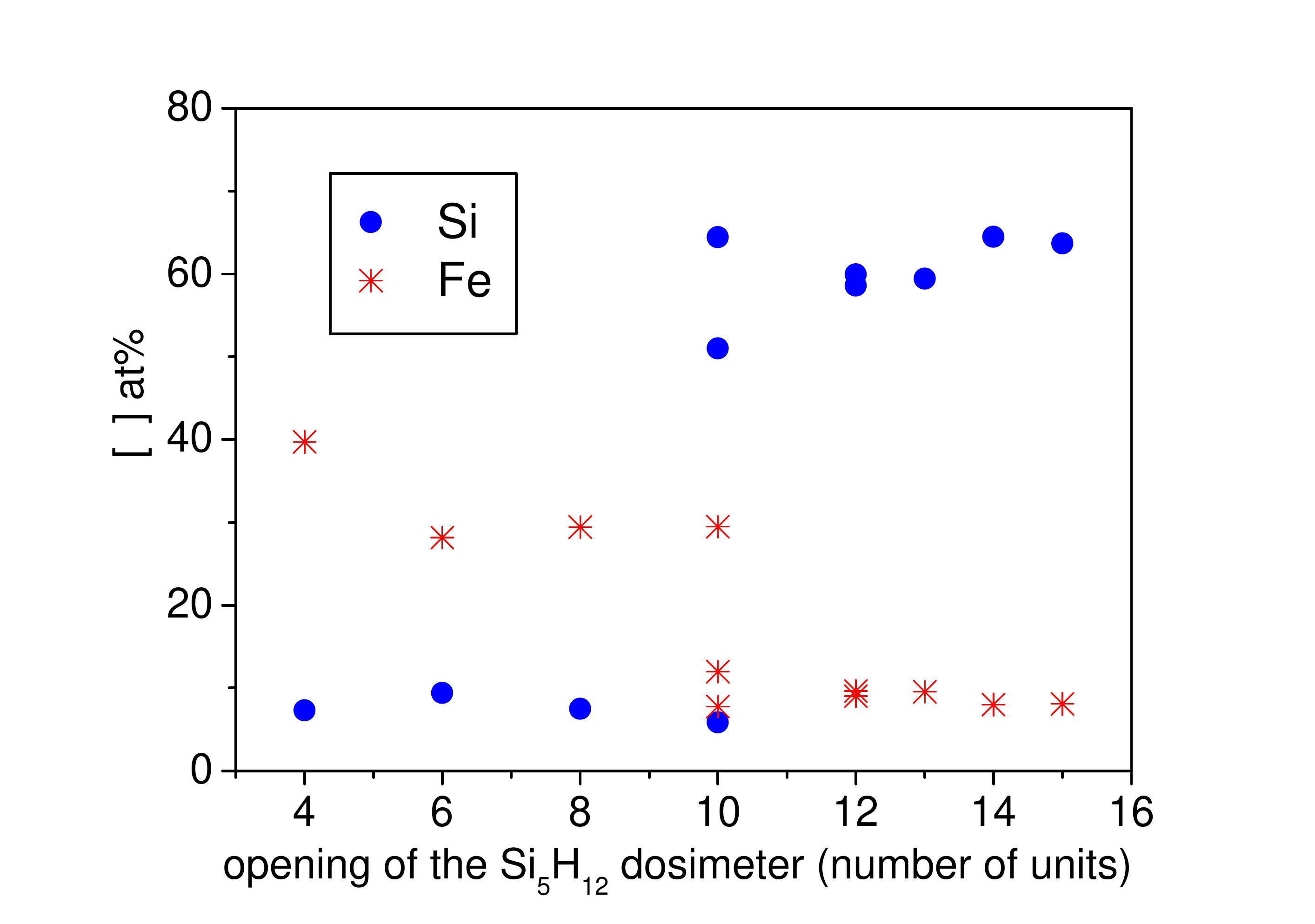}}\caption{EDX analysis of samples grown by the simultaneous use of $Fe(CO)_5$ and $Si_5H_{12}$. The $[Si]$ and $[Fe]$ concentrations are given as a function of the opening of the dosing valve containing the $Si_5H_{12}$ precursor. The base pressure of the SEM was about $4.7\cdot10^{-6}$~mbar. By opening the $Si_5H_{12}$ dosing valve the pressure increases approximatively linearly. The maximal opening of 15 units corresponds to a partial pressure of $5.0\cdot10^{-6}$~mbar. The sharp transition for $[Si]$ and $[Fe]$ indicates that $Fe(CO)_5$ and $Si_5H_{12}$ compete for adsorption sites on the surface.}\label{EDX}\end{figure}

In a first attempt to fabricate $\textit{Fe-Si}$ binary alloys we grew a series of samples by the simultaneous use of $Fe(CO)_5$ and $Si_5H_{12}$. These samples were fabricated with a fixed gas pressure of about $1\cdot10^{-5}$~mbar for $Fe(CO)_5$ and a variable one for $Si_5H_{12}$. The composition of a reference deposit prepared using only $Fe(CO)_5$ was 49~at$\%$ Fe, 43~at$\%$ O and 8~at$\%$ C. By increasing the pressure of $Si_5H_{12}$ the concentration of $[Fe]$ first decreased to 28~at$\%$ and then remained constant, (dosing valve settings 4 to 6 and 6 to 10) see Fig.~\ref{EDX}. In the same pressure regime $[Si]$ remained constant at about 5-10~at$\%$. At a $Si_5H_{12}$ pressure of ca. $4.9\cdot10^{-6}$~mbar (dosing valve setting 10) a sharp transition occurred, with a drop of $[Fe]$ to 6-8~at$\%$ and an increase of $[Si]$ to 60-65~at$\%$. This behavior strongly indicates that $Fe(CO)_5$ and $Si_5H_{12}$ compete for adsorption sites on the surface. According to the respective partial pressures of the two precursors, either bright $Fe$-rich deposits or dark $Si$-rich deposits are observed. Given the sharpness of the transition, it is very difficult to control the $[Fe]/[Si]$ ratio and, thus, to fabricate alloys with the desired stoichiometry. Therefore, as an alternative fabrication strategy we prepared $Fe/Si$ multilayers and we treated them to induce atomic species intermixing to obtain the desired binary compound.

During the fabrication of the multilayers, the pressure of $Fe(CO)_5$ and $Si_5H_{12}$ was $2\cdot10^{-5}$~mbar and $6\cdot10^{-6}$~mbar, respectively. $[Fe/Si]_2$ was fabricated by depositing alternatingly $Fe(CO)_5$ and $Si_5H_{12}$ for 30 seconds each. A waiting time of 10 seconds was used after the deposition of each layer. The composition of the $[Fe/Si]_2$ multilayers was calibrated with the support of a $[Fe/Si]_{15}$ multilayer, which is thick enough to avoid spurious contributions of the $Si_3N_4$ substrate during the energy dispersive x-ray (EDX) analysis. The composition of this multilayer was 13.6~at$\%$ C, 18.5~at$\%$ O, 31.4~at$\%$ Fe and 36.5~at$\%$ Si. $[Fe_3/Si]_2$ was fabricated by suitably tuning the fabrication parameters used for [Fe/Si]$_2$ , i.e., by increasing the deposition time of $Fe(CO)_5$ to 45 seconds and decreasing that of $Si_5H_{12}$ to 15 seconds. The thicknesses of these samples were not measured directly in order to avoid damaging the $Si_3N_4$ membranes. Nevertheless, an estimation based on samples grown on bulk $Si_3N_4$ indicates that the thickness of the multilayers was between 30~nm and 40~nm.

\begin{figure}[ht]\hspace{30mm}{\includegraphics[width=11cm]{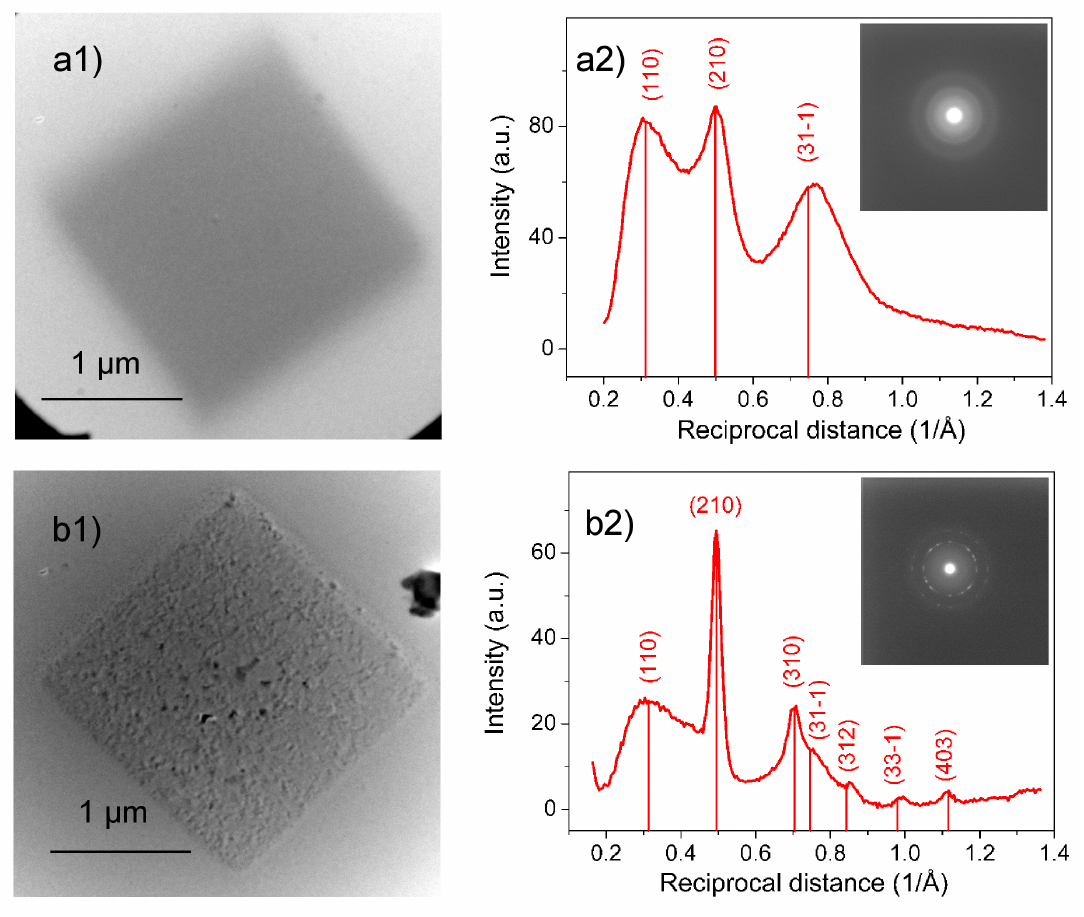}}\caption{a1) TEM image of a $[Fe/Si]_2$ multilayer prepared by using alternatingly $Fe(CO)_5$ and $Si_5H_{12}$. After growth the sample was electron-irradiated with a dose of 1~$\mu C/\mu m^2$ at 280~$^\circ$C. a2) Radial intensity of the diffraction pattern of the sample shown in a1). The three peaks are compatible with the $B20$ crystal structure of $FeSi$. In the inset the diffraction pattern is shown. b1) TEM image of a $[Fe/Si]_2$ multilayer grown as the one shown in a1). After growth and electron irradiation the sample was annealed in a furnace, see text for details. b2) Radial intensity of the diffraction pattern (see inset) of the sample shown in b1). The peaks are compatible with the $B20$ crystal structure of $FeSi$.}\label{FeSi_multilayer}\end{figure}

In Fig.~\ref{FeSi_multilayer}a1) and b1) we show TEM images of two $[Fe/Si]_2$ multilayers. After fabrication the samples were treated with electron beam irradiation at 280$^{\circ}$~C. Additionally, the multilayer of Fig.~\ref{FeSi_multilayer}b1) was annealed in a furnace at 600$^{\circ}$~C. In Fig.~\ref{FeSi_multilayer}a2) and b2) we provide the corresponding diffraction patterns. The radial intensity obtained by azimuthal integration of the diffraction pattern shows three diffuse peaks corresponding to the (110), (210) and (31$\overline{1}$) atomic planes of $FeSi$, see Fig.~~\ref{FeSi_multilayer}a2). We expect that a larger number of peaks might be obtained with a higher irradiation doses or by using membranes thinner than 100~nm, which leads to a stronger signal during sample imaging. A larger number of peaks compatible with the $B20$ phase of $FeSi$ is found in the sample which was annealed at 600$^{\circ}$~C, Fig.~\ref{FeSi_multilayer}b2). In particular, the (210) reflection is the most pronounced, as expected. The minor peaks (2kl) are not visible. The main peaks (3kl), i.e. (310), (31$\overline{1}$), (312), (33$\overline{1}$) are all present. The peaks (111) and (220), expected between the peaks (110) and (210) are not clearly visible in the diffraction pattern. Therefore their position is not indicated in Fig.~\ref{FeSi_multilayer}b2).

\begin{figure}[htp]\hspace{30mm}{\includegraphics[width=11cm]{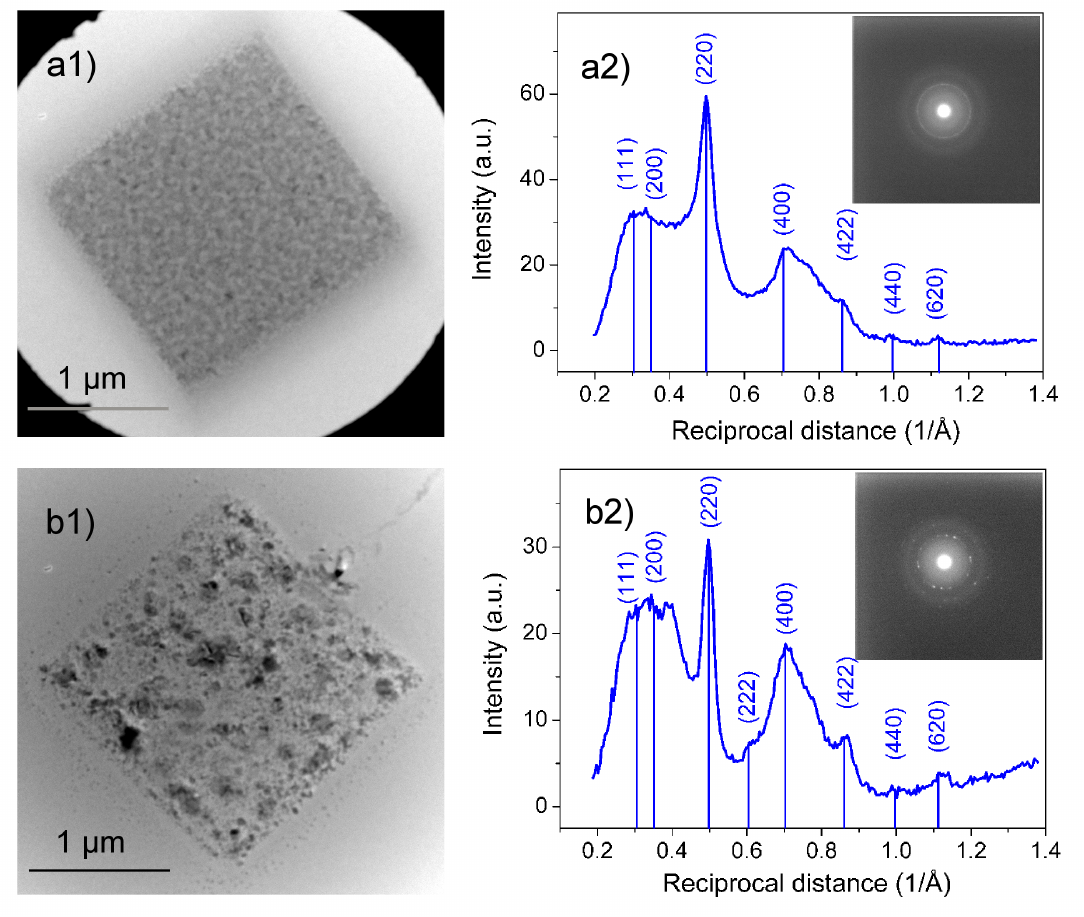}}\caption{a1) TEM image of a $[Fe_3/Si]_2$ multilayer prepared by using alternatingly $Fe(CO)_5$ and $Si_5H_{12}$ as precursors. After growth the sample was irradiated with the electron beam with a dose of 1~$\mu C/\mu m^2$ at 280~$^\circ$C. a2) Radial intensity of the diffraction pattern of the sample shown in a1). The peaks measured are compatible with the $D0_3$ crystal structure of the $Fe_3Si$ pseudo-Heusler compound. In the inset the diffraction pattern is shown. b1) TEM image of a $[Fe_3/Si]_2$ multilayer grown as the one shown in a1). After growth and electron irradiation the sample was in addition annealed in a furnace, see text for details. b2) Radial intensity of the diffraction pattern (see inset) of the sample shown in b1). The peaks are compatible with the $D0_3$ crystal structure of the $Fe_3Si$ pseudo-Heusler compound.}\label{Fe3Si_multilayer}\end{figure}

In Fig.~\ref{Fe3Si_multilayer} we report the results of a similar experiment, performed to fabricate $[Fe_3/Si]_2$ multilayers grown on $Si_3N_4$ membranes. In Fig.~\ref{Fe3Si_multilayer}a1) and b1) we provide TEM images of two multilayers treated with postgrowth electron irradiation at 280$^{\circ}$~C. Additionally, the multilayer shown in Fig.~\ref{Fe3Si_multilayer}b1) was annealed in a furnace at 600$^{\circ}$~C. The radial intensity of the diffraction pattern of the sample treated only with electron irradiation shows a number of peaks compatible with the $D0_3$ crystal structure of $Fe_3Si$, see Fig.~\ref{Fe3Si_multilayer}a2). A similar analysis was carried out for the diffraction pattern obtained from the sample annealed at 600$^{\circ}$~C, see Fig.~\ref{Fe3Si_multilayer}b2) . As for the previous case, this analysis shows a number of peaks compatible with the $D0_3$ crystal structure of $Fe_3Si$. In particular, the highest peak is the one corresponding to the (220) atomic planes of $Fe_3Si$, as expected. The next highest peaks, (400), (422), (440) and (620) are all visible. Among the minor peaks, only the (222) reflexion is visible in Fig.~\ref{Fe3Si_multilayer}b2). The position of the peaks (111) and (200) peaks is also indicated, despite of the fact that these peaks were not clearly detected.

Comparing the results obtained for $FeSi$ and $Fe_3Si$ we note that post-growth electron irradiation seems to induce better crystallization of the $D0_3$ phase, as is evident from the sharp diffraction peaks in Fig.~\ref{Fe3Si_multilayer}a2).

\begin{center}
\large\textbf{4. Discussion}\normalsize
\end{center}

Conventionally, the precursors used in FEBID are homonuclear complexes, i.e., they are molecules with a single metal atom species bound to, for example, organometallics, carbonyls or halides~\cite{utke08}. The deposits fabricated by means of these precursors are either amorphous or granular metals made of metallic nanoparticles embedded in a carbonaceous matrix. Depending on the metal atomic species and the atomic concentrations, either insulating, semiconducting, metallic or superconducting samples can be prepared~\cite{utke08,huth12,deteresa14}. Heteronuclear precursors, which contain more than one metal atom species, can be used to fabricate binary alloys, as we recently showed by preparing $CoFe$ nanostructures by employing the $HFeCo_3(CO)_{12}$ metal carbonyl~\cite{CoFe}. In this case, the metal stoichiometry of the precursor and of the deposits are identical. The use of heteronuclear precursors represents the most direct way to obtain binary alloys with a desired metal stoichiometry. If, on the other hand, a suitable precursor is not available, the fabrication of binary alloys can take place by the use of two homonuclear precursor gases, each containing one of the two required atomic species. In the last years, several precursors, like $Fe(CO)_5$, $Co_2(CO)_8$, $Si_5H_{12}$ and $(CH_3)_3CH_3C_5H_4Pt$ have been employed to fabricate $FePt$, $PtSi$ $CoPt$ and $CoSi$ alloys~\cite{che05,winhold11,CoPt,CoSi}. In these studies, in order to obtain the stoichiometry of interest, the precursors were mixed continuously by changing their relative flux. In the present investigation, the sharp transition between Fe-rich and Si-rich deposits shows that $Si_5H_{12}$ and $Fe(CO)_5$ compete for adsorption site during the fabrication process. Here we notice that understanding the reason for the sharp transition of Fig.~\ref{EDX} involves the solution of a system of Langmuir differential adsorption rate equations and the knowledge of the related adsorption, diffusion, desorption and decomposition parameters~\cite{bernau10,huth12,toth15}, which is beyond the aim of the present work. However, as an hint for future investigations, we observe that the residence time of $Si_5H_{12}$ in the SEM was rather long, since it was possible to grow Si-based deposits for some hours after having close the relative $Si_5H_{12}$ dosimeter. Therefore, we expect the sticking probability of $Si_5H_{12}$ to be much larger than the one of $Fe(CO)_5$ and, thus, the coverage probability of neopentasilane to be much higher than the one of iron pentacarbonyl. This picture is maintained until the flux of $Fe(CO)_5$ becomes dominant in comparison with the one of $Si_5H_{12}$. At that point, Fe-rich deposits are grown. Given the sharpness of the transition, deposits with continuously tuned stoichiometry cannot be easily obtained. In this case the fabrication can take place by growing multilayers by the alternating use of the two precursor gases and by inducing atomic species intermixing by means of a post-growth treatment. The stoichiometry of the deposits is then chosen by tuning the relative thickness of the layers.

In the literature there is a number of reports about the effect of electron beam irradiation on solids. For example, it was shown that electron beam irradiation can induce the formation of metal silicides on $Si$ substrates covered by metal films of $Ti$ and $Mo$~\cite{moore,suzuki}. Electron irradiation has been used to induce phase transformations in bulk and low dimensional systems~\cite{nagase,du}, and to tune the physical properties of graphene and carbon nanotubes~\cite{krasheninnikov}. In a series of recent papers we have started to investigate the effects of electron beam irradiation on the microstructure of FEBID samples~\cite{huth12}. For example, it was shown that during electron-beam irradiation the carbon-like matrix of $\textit{Pt-C}$ nanocomposites tends to graphitize, while the intergrain distance of the $Pt$ nanograins is reduced, which is due to carbon removal and to an increase of the diameter of the grains~\cite{PtC}. Furthermore, it was shown that electron-beam irradiation induces a room-temperature transformation from an amorphous phase into a crystalline one for $CoPt$ nanostructures~\cite{CoPt}. In that work it was speculated that the crystallization process starts with carbon removal, which facilitates solid-state diffusion of $Co$ and $Pt$ atoms. A similar mechanism is also evoked here, to explain the intermixing of $Fe$ and $Si$ atoms and the crystallization into the $B20$ and $D0_3$ phases of $FeSi$ and $Fe_3Si$, respectively. In all these experiments carbon removal is induced by an electron-beam-mediated reaction of carbon with residual gas molecules, like $H_2O$ and $O_2$, present in the microscope chamber~\cite{martin15}. Note that in Ref.~\cite{CoPt} atoms intermixing took place at room-temperature by using an irradiation dose of 8.64~$\mu C/\mu m^2$. In the present work, a home made heating stage was employed to increase the temperature of the samples to 280$^{\circ}$~C, which allowed to reduce the irradiation doses to 1~$\mu C/\mu m^2$. The temperature used is far away from those commonly used to induce crystallization of granular metals and thin films~\cite{che05,gazzadi11,wang03}. Therefore, we conclude that in our case the atoms intermixing is mainly due to the electron irradiation treatment.

The binary compounds fabricated in this work are interesting not only for their electronic and magnetic properties, with potential use in microelectronic and spintronic applications, but also because they are key materials for the synthesis of transition metal monosilicide and half-Heusler compounds. Indeed, the ability to fabricate $FeSi$ opens the route for the fabrication of silicides like $Fe_{1-x}Co_xSi$ or $Fe_{1-x}Mn_xSi$, which share with $FeSi$ the same cubic $B20$ crystal structure. Furthermore, $Fe_3Si$ is a pseudo-Heusler compound with $D0_3$ crystal structure, which is equivalent to the $L2_1$ crystal structure of the half-Heusler compounds~\cite{jing09}. Recently, metal silicides and Heusler nanowires have been synthesized and characterized by electrical transport measurements and electron holography~\cite{schmitt10,simon15}. By means of the FEBID technique nanostructures of any shape and dimension can be grown. In general, the microstructure of FEBID deposits is often granular, i.e., the samples are made of metallic nanoparticles embedded in a carbonaceous matrix. For the samples fabricated in this work the matrix, i.e., carbon and oxygen, amounts to about 30~at$\%$, as measured by EDX. Therefore it is expected that $FeSi$ and $Fe_3Si$ grains form nano-granular deposits with semiconducting and metallic behavior, respectively. For samples with lower metal content, which were not investigated in the present work, finite-size effects due to the granularity may appear. For example, in granular metals, an enhanced tunnel magnetoersistance effect (TMR) effect is measured at low temperatures, in the co-tunneling electrical transport regime~\cite{mitani98}. Furthermore, an additional enhancement of the TMR effect is expected in granular metals with high spin-polarization~\cite{inoue96}, such as granular Heusler compounds.

\begin{center}
\large\textbf{5 Conclusions}\normalsize
\end{center}

In this work we have fabricated $\textit{Fe-Si}$ binary alloys by focused electron beam induced deposition using $Fe(CO)_5$ and $Si_5H_ {12}$ as precursor gases. We have found that during deposition these precursors compete for adsorption sites on the surface of the substrate and, therefore, cannot be used simultaneously for the preparation of $\textit{Fe-Si}$ alloys over the complete mixing range. As alternative fabrication procedure we have grown $Fe/Si$ multilayers by the alternating use of $Fe(CO)_5$ and $Si_5H_ {12}$. After growth the multilayers have been treated by electron-beam irradiation and annealing to induce atomic intermixing and to obtain the desired binary compounds. In particular, we have prepared $[Fe/Si]_2$ and $[Fe_3/Si]_2$ multilayers and obtained $FeSi$ and $Fe_3Si$ binary compounds, as confirmed by TEM investigations. The ability to prepare these binary systems is the key for the synthesis of transition metal monosilicide and half-Heusler nanostructures, which share with $FeSi$ and $Fe_3Si$ either the same or an equivalent crystal structure.

\begin{center}
\large\textbf{6 Acknowledgments}\normalsize
\end{center}

Authors acknowledge A. Walters and L. Keller for technical support during the preliminary investigation of $\textit{Fe-Si}$ alloys. FP and MH acknowledge financial support by the Deutsche Forschungsgemeinschaft (DFG) under the project PO 1415/2-1 and HU 752/9-1, respectively. This work was carried out in the frame of the CELINA COST Action CM1301.

\newpage

\newpage

\begin{thebibliography}{99}
{\footnotesize\markboth{Bibliograbhy}{Bibliography}


\bibitem{okamoto00} \textit{Desk Handbook: Phase Diagrams for Binary Alloys}, edited by H. Okamoto (ASM International, 2000).

\bibitem{liang11} Y. F. Liang, S. L. Shang, J. Wang, F. Ye, J. P. Lin, G. L. Chen, and Z. K. Liu, Intermetallics \textbf{19}, 1374 (2011).


\bibitem{leong97} D. Leong, M. Harry, K. J. Reeson, and K. P. Homewood, Nature \textbf{387}, 686 (1997).


\bibitem{schlesinger93} Z. Schlesinger, Z. Fisk, Hai-Tao Zhang, M. B. Maple, J. F. DiTusa and G. Aeppli, Phys. Rev. Lett. \textbf{71}, 1748 (1993).


\bibitem{herfort03} J. Herfort, H. P. Sch\"{o}nherr, and K. H. Ploog, Appl. Phys. Lett. \textbf{83}, 3912 (2003).


\bibitem{veetil10} M. K. Kolel-Veetil and T. M. Keller, Materials \textbf{3}, 1049 (2010).



\bibitem{degroot88}  R. A. de Groot, F. M. Mueller, P. G. van Engen, K. H. J. Buschow, Phys. Rev. Lett. \textbf{50}, 2024 (1988).

\bibitem{felser07} C. Felser,  G. H. Fecher, and B. Balke, Angew. Chem. Int. Ed. \textbf{46}, 668 (2007).




\bibitem{walser76} R. M. Walser and R. W. Bene, Appl. Phys. Lett. \textbf{28}, 624 (1976).


\bibitem{gaffet93} E. Gaffet, N. Malhoureux, and M. Abdellaoui, J. Alloys Compd. \textbf{194}, 339 (1993).


\bibitem{dormans91} G. J. M. Dormans, J. Cryst. Growth \textbf{108}, 806 (1991).










\bibitem{zybill94} C. Zybill, H. Handwerker, and H. Friedrich, Adv.Organomet. Chem. \textbf{36}, 229 (1994).


\bibitem{schmitt10} A. L. Schmitt, J. M. Higgins, J. R. Szczech, and S. Jin, J. Mater. Chem. \textbf{20}, 223 (2010).


\bibitem{che05} R. C. Che, M. Takeguchi, M. Shimojo, W. Zhang, and K. Furuya, Appl. Phys. Lett. \textbf{87}, 223109 (2005).


\bibitem{winhold11} M. Winhold, C. H. Schwalb, F. Porrati, R. Sachser, A. S. Frangakis, B. K\"{a}mpken, A. Terfort, N. Auner and M. Huth, ACS Nano \textbf{5}, 9675 (2011).


\bibitem{CoPt} F. Porrati, E. Begun, M. Winhold, C. H. Schwalb, R. Sachser, A. S. Frangakis, and M. Huth, Nanotechnology \textbf{23}, 185702 (2012).


\bibitem{CoSi} F. Porrati, B. K\"{a}mpken, A. Terfort, and M. Huth, J. App. Phys. \textbf{113}, 053707 (2013).







\bibitem{utke08} I. Utke, P. Hoffmann, and J. Melngailis, J. Vac. Sci. Technol. B \textbf{26}, 1197 (2008).


\bibitem{huth12} M. Huth, F. Porrati, C. Schwalb, M. Winhold, R. Sachser, M. Dukic, J. Adams, and G. Fantner, Beilstein J. Nanotechnol. \textbf{3}, 597 (2012).


\bibitem{CoFe} F. Porrati, M. Pohlit, J. M\"{u}ller, S. Barth, F. Biegger, C. Gspan, H. Plank and M. Huth, Nanotechnology \textbf{26}, 475701 (2015).






\bibitem{lukas} L. Keller, Master thesis, Goethe-University, Frankfurt (2014).



\bibitem{deteresa14}  J. M. De Teresa and A. Fernandez-Pacheco \textit{Appl. Phys. A} \textbf{117}, 1645 (2014).



\bibitem{bernau10} L. Bernau, M. Gabureac, R. Erni, and I. Utke \textit{Angew. Chem., Int. Ed.} \textbf{49}, 8880 (2010).


\bibitem{toth15} M. Toth, C. Lobo, V. Friedli, A. Szkudlarek and I. Utke \textit{Beilstein J. Nanotechnol.} \textbf{6}, 1518 (2015)


\bibitem{moore} C. A. Moore, J. J. Rocca, G. J. Collins, P. E. Russell, and J. D. Geller,
Appl. Phys. Lett. \textbf{45}, 169 (1984).


\bibitem{suzuki} S. Suzuki, Y. Ohkubo, F. Matsuoka, and T. Itoh, Appl. Phys. Lett. \textbf{42}, 797 (1983).







\bibitem{nagase} T. Nagase, Y. Umakoshi, and N. Sumida , Mater. Sci. Eng. A
\textbf{323}, 218 (2002).

\bibitem{du} X. Du, M. Takeguchi, M. Tanaka, and K. Furuya, Appl.
Phys. Lett. \textbf{82}, 1108 (2003).

\bibitem{krasheninnikov} A. V. Krasheninnikov and K. Nordlund, J. Appl. Phys.
\textbf{107}, 071301 (2010).




\bibitem{PtC} F. Porrati, R. Sachser, C. H. Schwalb, A. S. Frangakis, and M. Huth,
J. Appl. Phys. \textbf{109}, 063715 (2011).


\bibitem{martin15} A. A. Martin, S. Randolph, A. Botman, M. Toth, and I.
Aharonovich, Sci. Rep. \textbf{5}, 8958 (2015).


\bibitem{gazzadi11} G. C. Gazzadi, H. Mulders, P. Trompenaars, A. Ghirri, M. Affronte, V. Grillo, and S. Frabboni, J. Phys. Chem. C \textbf{115}, 19606 (2011).


\bibitem{wang03} A. Wang, T. Li, Y. Zhou, H. Jiang, and W. Zheng, Thin Solid Films \textbf{445}, 127 (2003).



\bibitem{jing09} Y. Jing, Y. Xu, and J.-P. Wang, J. Appl. Phys. \textbf{105}, 07B520 (2009).


\bibitem{simon15} P. Simon, D. Wolf, C. Wang, A. A. Levin, A. Lubk, S. Sturm, H. Lichte, G. H. Fecher, and C. Felser, Nano Lett. \textbf{16}, 114 (2016).


\bibitem{mitani98} S. Mitani, S. Takahashi, K. Takanashi, K. Yakushiji, S. Maekawa,
and H. Fujimori, Phys. Rev. Lett. \textbf{81}, 2799 (1998).


\bibitem{inoue96} J. Inoue and S. Maekawa, Phys. Rev. B \textbf{53}, R11927 (1996).




}


\end{thebibliography}
\end{document}